\documentclass[pre,a4paper,
twocolumn,amsmath,superscriptaddress,showpacs]{revtex4}
\usepackage{graphicx}
\usepackage{bm}
\newcommand{\bee}{\begin{eqnarray}}
\newcommand{\eee}{\end{eqnarray}}
\newcommand{\ba}{\begin{array}}
\newcommand{\ea}{\end{array}}

\usepackage{times}
\usepackage{accents}

\begin{document}
\newcommand{\bZ}{{\mathbf Z}}
\newcommand{\m}{\mbox{\boldmath $\mu$}}
\newcommand{\aaa}{\alpha}
\newcommand{\bb}{\beta}
\title{Short-time dynamics and high-frequency rheology of core-shell suspensions with thin-shells}

\author{Bogdan Cichocki}
 \affiliation{Institute of Theoretical Physics, Faculty of Physics, University of Warsaw, Ho\.za 69,
  00-681 Warsaw, Poland}

\author{Maria L. Ekiel-Je\.zewska}
\email{mekiel@ippt.pan.pl}
 \affiliation{Institute of Fundamental Technological Research,
             Polish Academy of Sciences, Pawi\'nskiego 5B, 02-106 Warsaw, Poland}

\author{Eligiusz Wajnryb}
 \affiliation{Institute of Fundamental Technological Research,
             Polish Academy of Sciences, Pawi\'nskiego 5B, 02-106 Warsaw, Poland}

\date{\today}

\begin{abstract}
Short-time dynamics and high-frequency rheology for suspensions of non-overlapping core-shell particles with thin shells were analysed. In the thin-shell limit, the single-particle scattering coefficients were derived and shown to define a unique effective radius. This result was used to justify theoretically (in the thin-shell limit) the accuracy of the 
annulus approximation with the inner radius equal to the effective hydrodynamic radius of the core-shell particle.  
The two-particle virial expansion of the translational \& rotational self-diffusion, sedimentation and viscosity was performed. The virial coefficients were evaluated and shown to be accurately approximated by the effective annulus model, in contrast to the imprecise effective hard sphere model. 
\end{abstract}


\maketitle

\section{Introduction}\label{int}
Recently, there has been a growing interest in micro and nanogels, and other permeable particles, which can be used to carry drugs or proteins \cite{Kasper:98}-\cite{Fuchs:12}. 
For such systems, density of polymer segments inside the core region is frequently much higher than in the outer part, and therefore they are often approximated as core-shell particles.

\begin{figure}[h]
\includegraphics[width=3.5cm]{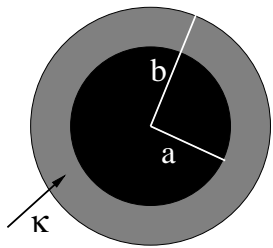}
\caption{The core-shell particle: the solid core of radius $a$ is surrounded by the uniformly permeable shell of radii $a$, $b$ and the hydrodynamic penetration depth $\kappa^{-1}$.}\label{fi1}
\end{figure}

A core-shell particle consists of a solid core of radius $a$, and a surrounding permeable shell, with the inner and outer radii, $a$ and $b$, respectively (see Fig.~\ref{fi1}). 
The porous medium inside the shell is characterized by 
the uniform hydrodynamic
penetration depth $\kappa^{-1}$. It is assumed that the particles do not overlap, i.e. their centers cannot come closer to each other than $2b$.

In a recent paper~\cite{Abade:12}, the precise  multipole method was applied to extensively analyze dynamics and rheology of core shell particles. High-frequency viscosity, short-time translational and rotational diffusion, and sedimentation coefficient have been evaluated numerically for a very wide range of permeabilities of both wide and thin shells, with volume fractions up to $\phi=0.45$. 
It was concluded that for wide shells with a low permeability ($\kappa (b-a)\gtrsim 5$), the core is practically invisible and particles behave as uniformly permeable.  

In the oposite limit of thin shells, it was shown that the core-shell particles cannot be well-approximated by hard spheres of radius $b$, even for a tiny, hardly permeable shell. However, the dynamics of concentrated core-shell systems is well reproduced, if a thin permeable shell is replaced by the shell of the pure fluid (the so-called annulus model~\cite{Felderhof_Cichocki:91}). The difference between the hard spheres and annulus particles of the same radii $b$ is very large; for example, 20-35\% for the diffusion and sedimentation, and 60\% for the viscosity, if $(b-a)/a=0.05$, $\kappa a=10$, and $\phi=0.45$. 
 The shell influence cannot be neglected, and the essential reason is that it prevents particles from overlapping.

This result indicates that a more detailed analysis of the thin-shell limit is needed. 
For dilute systems of non-interacting core-shell particles, 
the single-particle translational \& rotational diffusion coefficients and intrinsic viscosity determine the corresponding hydrodynamic radii $a_{\text{eff}}$~\cite{Masliyah:87,ChenYe:00,Zackrisson_Bergenholtz:03,Cichocki_Felderhof:09}, which, in general, are  only tenuously connected to the geometrical radius $a$ of the core-shell particle~\cite{Zackrisson_Bergenholtz:03}. A thin-shell analysis of such systems has been recently performed in Ref.~\cite{Zackrisson_Bergenholtz:03}. The leading terms of $a_{\text{eff}}$ have been given and shown to be the same for the intrinsic viscosity and the translational diffusion.

In this paper, we extend the idea of Ref.~\cite{Zackrisson_Bergenholtz:03}, and we investigate the thin-shell limit for concentrated systems of the core-shell particles. We use 
 the general concept of the hydrodynamic radius model (HRM) \cite{Abade_JCP:2010}, where the effective radius approximation is restricted only to the hydrodynamic interaction, with no modification of direct interactions. In this work, the direct interactions between core-shell particles are specified by the no-overlap radius $b$. In this case, the HRM model is equivalent to the effective annulus model with the no-overlap radius of the outer shell $b$ and the radius of the inner shell equal to the hydrodynamic radius $a_{\text{eff}}$ of 
the core-shell particle. We will show that the effective annulus approximation is much more precise than the standard model of a hard-core particle with the effective radius $a_{\text{eff}}$.

The paper is organized as follows. Sec.~\ref{II} contains theoretical description of the core-shell suspensions.
Sec.~\ref{main} is focused on the hydrodynamics of the single core-shell particles with thin shells. In Sec.~\ref{AA}, direct (no-overlap) interactions between core-shell particles are described within the annulus model. In Sec.~\ref{IV}, the thin-shell limit of the two-particle virial coefficients is explicitly carried out for the translational \& rotational self-diffusion, sedimentation and effective viscosity, and accuracy of the effective annulus model is examined. The conclusions are presented in Sec.~\ref{con}.

\section{The core-shell suspensions}\label{II}
We consider a fluid of shear viscosity $\eta_0$, with randomly distributed identical core-shell particles~\cite{Cichocki_Felderhof:09} defined in Sec.~\ref{int} and shown in Fig.~\ref{fi1}. 

Geometry of the particle is characterized by the relative thickness of the porous shell,
\begin{equation}
\epsilon =\frac{b-a}{a}.\label{thin}
\end{equation}
The effect of the particle permeability is described by the ratio $x$ of the particle radius $a$ to the hydrodynamic screening length $\kappa^{-1}$ of the uniformly porous material inside the particle, i.e.
\bee
x &=& \kappa a.
\eee 
All the distances are normalized by the core radius $a$.

The fluid flow is characterized by the low-Reynolds-number, Re$<<$1. 
Outside the particles, the fluid velocity ${\bm v}$ and pressure $p$ satisfy the Stokes equations \cite{Kim-Karrila:1991,Happel-Brenner:1986},
\bee
    \eta_0\;\! {\bm \nabla}^2 {\bm v}({\bm r})  - {\bm \nabla} p({\bm r})  &=& 0 \nonumber \\
    {\bm \nabla} \cdot {\bm v}({\bm r}) &=& 0,
\eee
and inside the particle shells, by the Brinkman-Debye-B\"uche (BDB) equations \cite{Brinkman:47,DebyeBueche:48},
\bee
\label{eq:DBB}
    \eta_0\;\! {\bm \nabla}^2 {\bm v}({\bm r}) - \eta_0\;\! \kappa^2 \left[{\bm v({\bm r})} -
    {\bm u}_i({\bm r}) \right] - {\bm \nabla} p({\bm r})  &=& 0 \nonumber \\
    {\bm \nabla} \cdot {\bm v}({\bm r}) &=& 0.
\eee
The core and the skeleton of the particle $i$, centered at ${\bf r}_i$,   move rigidly with the local  velocity 
${\bf u}_i({\bf r}) = {\bf U}_i + {\bm{\Omega}}_i \times \left( {\bf r} - {\bf r}_i \right)$,
determined by their translational and rotational velocities ${\bf U}_i$ and ${\bm{\Omega}}_i$, respectively. The fluid velocity and the 
stress are continuous across the particle surface, i.e. at $|{\bf r}- {\bf r}_i|=b$. The fluid sticks to the core surface; that is, 
${\bm v}=0$ at $|{\bf r}- {\bf r}_i|=a$.

The coupled Stokes-BDB problem defined above is solved by the multipole expansion \cite{CFHWB:94,E-JW:2009}. Many-particle mobility coefficients are evaluated and averaged over the statistical equilibrium ensemble of configurations, resulting in the short-time self-diffusion, sedimentation and high-frequency viscosity coefficients, see e.g. Refs.~\cite{m1,m2,m3,m4,Abade:12} for the general theoretical scheme. The important issue discussed in this work is that the N-particle mobility coefficients $\m_{ij}(1...N)$
are 
determined in terms of the Green operators ${\bf G}(mn)$ (which depend on the distance between particles $m$ and $n$), and the single-particle friction operators, ${\bf Z}_0(k)$ and $\hat{\bf Z}_0(k)$ (which depend on the internal structure and size of a particle $k$ \cite{Cichocki_Felderhof_Schmitz:88}). This relation can be written as a scattering series,
\bee
&&\m_{ij}(1...N)\!\!\!=\!\!\!\delta_{ij}\,\m_0(i) + \m_0(i)\, {\bf Z}_0(i) \sum^N_{k=1}\left[\delta_{ik}\right.\nonumber \\
&&\left.-(\!1\!\!-\!\delta_{ik}\!)\,{\bf G}(ik)\,
\hat{\bf Z}_0(k)\!+...\right] (\!1\!\!-\!\delta_{kj}\!)\, {\bf G}(kj)\, {\bf Z}_0(j)\, \m_0(j).\nonumber  \\
\label{scm}
\eee
or in short as
\bee
\m =\m_0 + \m_0 \;\bZ_0 \;\frac{1}{{\bf 1}+{\bf G}\hat{\bZ}_0}\;{\bf G}\; 
\bZ_0 \;\m_0 , \label{scattmobility}
\eee
with $\m_0$ denoting the single-particle mobility operator (determined by the inverse of ${\bf Z}_0$).
The single-particle friction operators, ${\bf Z}_0$ and $\hat{\bf Z}_0$, determine the hydrodynamic force density exerted by a given ambient flow on a motionless and a freely moving particle, respectively. 
The matrix elements of these operators are
the single particle scattering coefficients $A_{l\sigma}$, with $l\!=\!1,2,3,4,...$ and $\sigma\!=\!1,2,3$ ~\cite{Cichocki_Felderhof_Schmitz:88}, see Appendix~\ref{ap1} for the explicit relation. If the ambient fluid flow is decomposed into the multipole components~\cite{Lamb}, labeled by $\sigma,\,l$ and $m\!=\!0,...,\pm l$, then $A_{l\sigma}$ 
determine 
the corresponding multipoles of the fluid velocity, reflected (scattered) by a particle 
 immersed in a given ambient flow. This is why $A_{l\sigma}$ are called ``scattering coefficients''\cite{Cichocki_Felderhof:09,Cichocki_Felderhof_Schmitz:88}. In the multipole approach, differences in the internal structure of particles (e.g. solid, liquid, gas, porous, core-shell, stick-slip), are fully accounted by different scattering coefficients. The other parts of the 
multipole algorithm need not to be changed.
In particular, the scattering coefficeints $A_{10},\;A_{11}$ and $A_{20}$ determine the single-particle translational, $D_{0}^{t}$, and rotational, $D_{0}^{r}$, diffusion coefficients and the intrinsic viscosity, $\bar{\left[\eta \right]}$, by the following relations,
\bee
\bar{\left[\eta \right]} &=& \frac{3\mu _{0}^{d}}{8 \pi \eta_0 a^3},\label{invis}\\
D_{0}^{t}&=& k_{B}T\mu _{0}^{t},\label{D0t}\\
D_{0}^{r}&=& k_{B}T\mu _{0}^{r},
\eee 
with the corresponding single-particle mobility coefficients,
\bee
\mu _{0}^{d}&=&\frac{8\pi \eta_0 A_{20} }{3},\label{mud1}\\
\mu _{0}^{t}&=&\frac{1}{4\pi \eta_0 A_{10}},\label{mut1}\\
\mu _{0}^{r}&=&\frac{1}{8\pi \eta_0 A_{11}}.\label{mur1}
\eee

 For core-shell particles, the dependence of all the scattering coefficeints $A_{l\sigma}$ on $(x,a,b)$ have been determined analytically in Ref.~\cite{Cichocki_Felderhof:09}.

\section{Single core-shell particles with thin shells}\label{main}

In this section, we perform asymptotic expansion of the core-shell model in the thin-shell limit, i.e. when
\bee a, \kappa=\mbox{const} \;\;\mbox{ and }\;\; \epsilon \rightarrow 0.\label{thin-shell} \eee

The asymptotic expansion of all the single-particle scattering coefficients $A_{l\sigma}(x,a,b)$, derived in Ref.~\cite{Cichocki_Felderhof:09},  in the thin-shell limit \eqref{thin-shell}, is the following, 
\bee
A_{l\sigma }(x,a,b)&=&A_{l\sigma }^{hs}(a)[1+\aaa_{l\sigma }x^{2}\epsilon
^{3}/3+...]
,\label{a1}\\
B_{l2}(x,a,b)&=&B_{l2}^{hs}(a)[1+\bb_{l2}x^{2}\epsilon
^{3}/3+...]
,\label{a2}
\eee
where $l=1,2,...$, $\sigma=0,1,2$, 
\bee
\aaa_{l0} &=& {2l-1},\\
\aaa_{l1} &=& \aaa_{l2} = {2l+1},\\
\bb_{l2} &=& {2l+3},
\eee
and the terms omitted in Eqs.~\eqref{a1}-\eqref{a2}  scale as $\sim x^2\epsilon ^{4}$.

Above,  $A_{l\sigma }^{hs}(a)$ and $B_{l2}^{hs}(a)$ denote the scattering coefficients for a hard sphere of radius $a$ \cite{Cichocki_Felderhof_Schmitz:88}, listed in Appendix \ref{ap1}.
Taking into account that
\bee
A_{l0}^{hs}(a) &\sim& a^{2l-1}=a^{\aaa_{l0}} , \\
A_{l1}^{hs}(a), A_{l2}^{hs}(a) &\sim& a^{2l+1}=a^{\aaa_{l1}}=a^{\aaa_{l2}}, \\
B_{l2}^{hs}(a) &\sim& a^{2l+3}=a^{\aaa_{l2}},
\eee
and using the relation,
\bee
a^{k}(1+k \psi) \stackrel{\psi \rightarrow 0}{\approx}[a (1+\psi)]^{k},
\eee
with $\psi=x^2\epsilon^3/3$ and $k=\alpha_{l\sigma},\beta_{l\sigma}$, 
we can combine the leading terms of the asymptotic expansion \eqref{a1}-\eqref{a2}, and write it in the following compact form, 
\bee
A_{l\sigma }(x,a,b)&=&A_{l\sigma }^{hs}(a_{\text{eff}})[1+{\cal O}(\epsilon ^{4})],\\
B_{l2}(x,a,b)&=&B_{l2}^{hs}(a_{\text{eff}})[1+{\cal O}(\epsilon ^{4})],
\eee
with the universal $a_{\text{eff}}$ for all the multipole indices,
\bee
a_{\text{eff}}/a=1+\frac{1}{3} x^2 \epsilon^3 + {\cal O}(\epsilon ^{4}),\label{thesame}
\eee
and the ${\cal O}(\epsilon ^{4})$ terms $\sim x^2\epsilon ^{4}$.
The above relations show that in the leading order of the thin-shell expansion, the core-shell scattering coefficients are equal to the scattering coefficients  $A_{l\sigma }^{hs}(a_{\text{eff}})$ and $B_{l2}^{hs}(a_{\text{eff}})$ of a hard sphere with the effective radius $a_{\text{eff}}$.
This result is the important generalization of the thin-shell analysis performed in Ref. \cite{Zackrisson_Bergenholtz:03} for the intrinsic viscosity and translational self-diffusion. 

In Ref.~\cite{Zackrisson_Bergenholtz:03} it was shown that the hydrodynamic radii, defined by the intrinsic viscosity, 
\bee
a_{\text{eff},\eta} &=& \left( \frac{2}{5}A_{20}\right)^{\!\!1/3}\label{r1},
\eee
 and the translational diffusion coefficient of single core-shell particles~\cite{Masliyah:87}, 
\bee
a_{\text{eff},t} 
= \frac{2}{3} A_{10},\label{r2}
\eee
have the same leading terms \eqref{thesame} in the thin-shell expansion.

We find out that all the scattering coefficients $A_{l\sigma}(x,a,b)$, up to terms $\sim (x^2\epsilon^3)$,  are determined by the same effective radius \eqref{thesame}. In particular, also
 the hydrodynamic radius defined by the single-particle rotational diffusion coefficient \cite{ChenYe:00},
\bee
a_{\text{eff},r} &=& A_{11}^{1/3}.\label{r3}
\eee

As discussed in the previous section, the many-body hydrodynamic interactions depend on the particle internal structure only through the scattering coeffcients. 
The result is that in the thin-shell expansion~\eqref{thin-shell} the hydrodynamic interactions between core-shell particles consist of the dominant contribution from the hard spheres with the effective radius, plus a small correction which depends on a single parameter $x^2\epsilon^4$.

In the next section, the single-particle scattering coefficients will be used to construct a model for the concentrated suspensions of core-shell particles.

\section{Annulus model}\label{AA}
\begin{figure}[h]
\includegraphics[width=2.75cm]{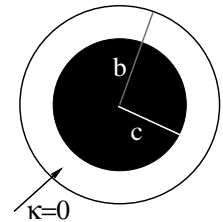}
\caption{The annulus particle: the solid core of radius $c$ 
is surrounded by a completely permeable ($\kappa\!=\!0$) shell with the no-overlap radius~$b$.}
\label{fig2}
\end{figure}

For semi-dilute and concentrated suspensions, 
direct interactions between particles have to be taken into account. 
In the annulus model~\cite{Felderhof_Cichocki:91}, 
these interactions are taken into account 
by introducing a non-overlap radius $b$, 
which surrounds a 
hard sphere of radius $c$, which is smaller than $b$. 
The shell of the relative thickness $\delta$,
\bee
\delta=\frac{b-c}{c},\label{del}
\eee
is filled with the same fluid as outside the annulus particle, see Fig.~\ref{fig2}. The annulus model can in general be applied for particles of a different internal structure and a different type of direct interactions. 

\begin{figure}[h]
\includegraphics[width=7cm]{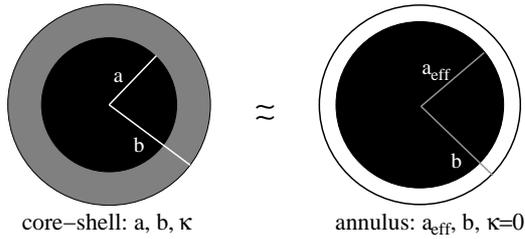}
\caption{In the dynamics of concentrated suspensions, a core-shell particle with a thin shell is well-approximated by the annulus model with the hydrodynamic inner radius $c=a_{\text{eff}}$.}\label{fi2}
\end{figure}

Transport coefficients of semi-dilute and concentrated  core-shell suspensions, such as 
the translational \& rotational self-diffusion, sedimentation and viscosity coefficients, are evaluated as statistical averages of the corresponding many-particle mobilities. 
The leading contributions to the moblilities of the core-shell particles with the outer and inner radii $b$ and $a$ (see Fig.~\ref{fi1}) come from the  hydrodynamic intractions of the hard-spheres with effective radii $a_{\text{eff}}$, larger than $a$ and smaller than $b$. On the other hand, the averaging 
takes into account the no-overlap condition - the particle centers have to be separated by a distance at least 
twice as large as the outer shell radius $b$. Therefore, the dynamics of many core-shell particles is approximated within the (effective) annulus model with $c=a_{\text{eff}}$ and the non-overlap radius $b$ the same as for the core-shell particle,  
see Fig.~\ref{fi2}.

The corresponding effective hydrodynamic shell thickness,
\bee
\epsilon_{\text{eff}}=\frac{b-a_{\text{eff}}}{a_{\text{eff}}}.\label{epse}
\eee

\section{Example of the thin-shell analysis: viral expansion of the transport coefficients}\label{IV}
In this Section, we will explicitly perform the thin-shell expansion for the two-particle virial contribution to the translational and rotatinal self-diffusion, sedimentation and high-frequency viscosity, and compare the results with the effective annulus approximation.  
\subsection{Virial expansion}\label{vi}
The virial expansion of the transport coefficients for suspensions of core-shell particles is carried out using the inner radius $a$ 
as the reference one, i.e. we
expand in powers of the volume fraction,
\begin{equation}
\phi_a =\frac{4}{3}\pi na^{3}.\label{phia}
\end{equation}

The short-time translational and rotational self-diffusion coefficients, $D_t$ and $D_r$, the sedimentation coefficient $K$, and the high-frequency viscosity $\eta_{\infty}$ have the following virial expansion,
\bee
D_{t}/D_{0}^{t}&=&1+\bar{\lambda} _{t}\phi_a +{\cal O}(\phi_a^2),\label{virS}\\
D_{r}/D_{0}^{r}&=&1+\bar{\lambda} _{r}\phi_a +{\cal O}(\phi_a^2),\\
K&=&1+\bar{\lambda} _{K}\phi_a +{\cal O}(\phi_a^2),\\
\eta_{\infty}/\eta_0 &=&1+\bar{[\eta]}\phi_a +\bar{\lambda}_{\eta}\phi_a^2 + {\cal O}(\phi_a^3),\label{vireta}
\eee
with the single core-shell particle translational \& rotational diffusion coefficients, and the intrinsic viscosity defined in Eqs.~\eqref{D0t}-\eqref{mur1}.    The bar over the virial coefficients reminds that they correspond to the expansion with respect to volume fraction based on the inner radius 
(rather than the outer one).

The explicit theoretical expressions for the two-particle virial 
coefficients $\bar{\lambda}\! \equiv \!\bar{\lambda}_t, \bar{\lambda}_K, \bar{\lambda}_r$ and $\bar{\lambda}_{\eta}$ are derived in Appendix~\ref{A}. In Sec.~\ref{results}, they are used to evaluate $\bar{\lambda}(x,\epsilon)$ numerically in a wide range of the parameters.

\subsection{Annulus model}
An annulus particle, introduced in Sec.~\ref{AA} and illustrated in Fig~\ref{fig2}, is described by two parameters: the hard core radius $c$ and the no-overlap radius $b$. The corresponding shell thickness $\delta$ is defined in Eq.~\eqref{del}. With the use of the volume fraction $\phi_c=4\pi n c^3/3$, the annulus two-particle virial coefficients $\bar{\lambda}^A(\delta)$ of the translational and rotational self-diffusion, sedimentation and viscosity are defined by the analogs of Eqs.~\eqref{virS}-\eqref{vireta} with $\phi_a \rightarrow \phi_c$, and the single core-shell particle coefficients 
\eqref{D0t}-\eqref{mur1} replaced by the corresponding values for a hard sphere with radius $c$.
The two-particle virial coefficients  $\bar{\lambda}^A(\delta)$ are evaluated and listed in table \ref{tab1}.

\begin{figure*}[ht]
\includegraphics[width=8.6cm]{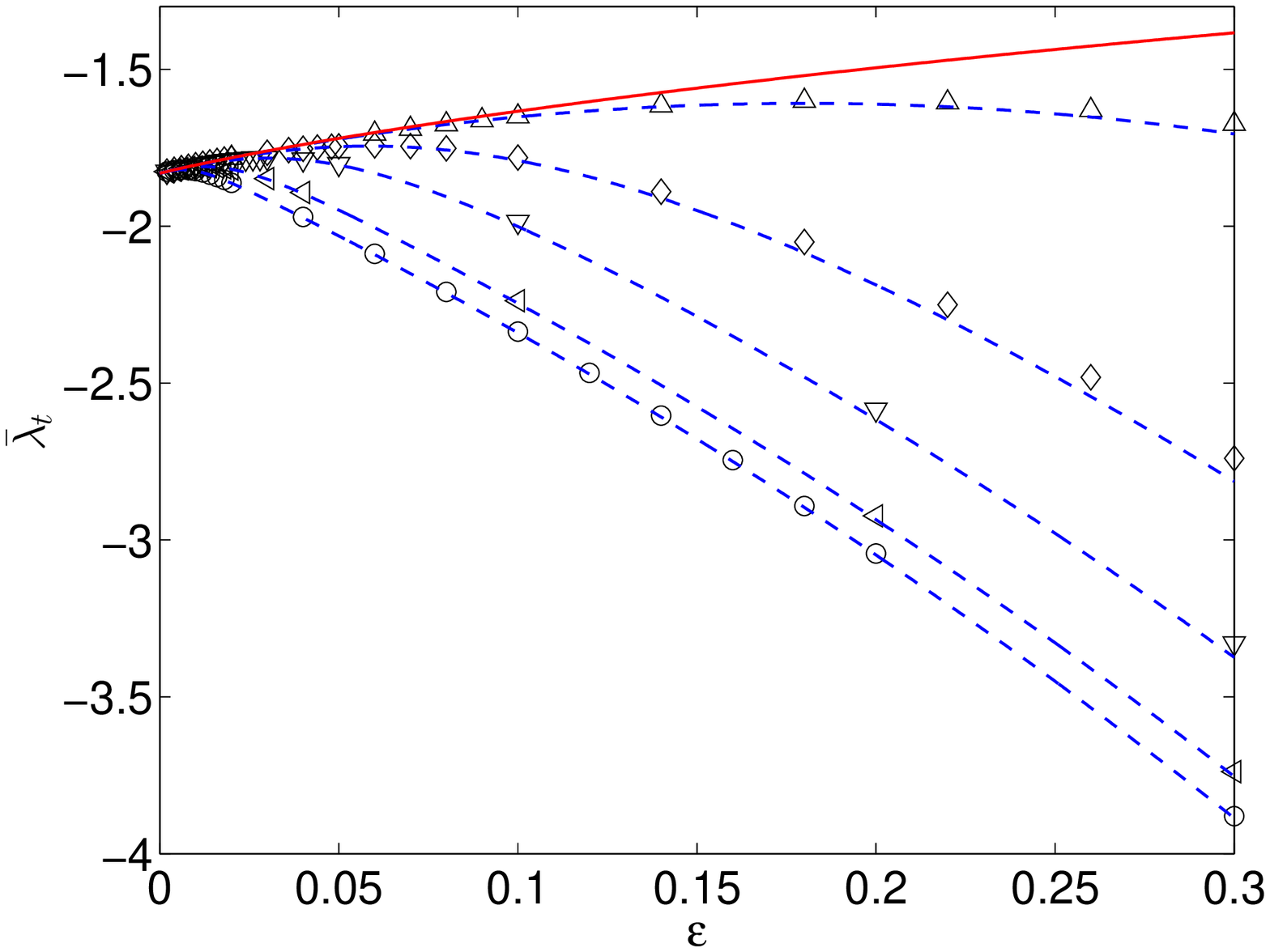} \includegraphics[width=8.6cm]{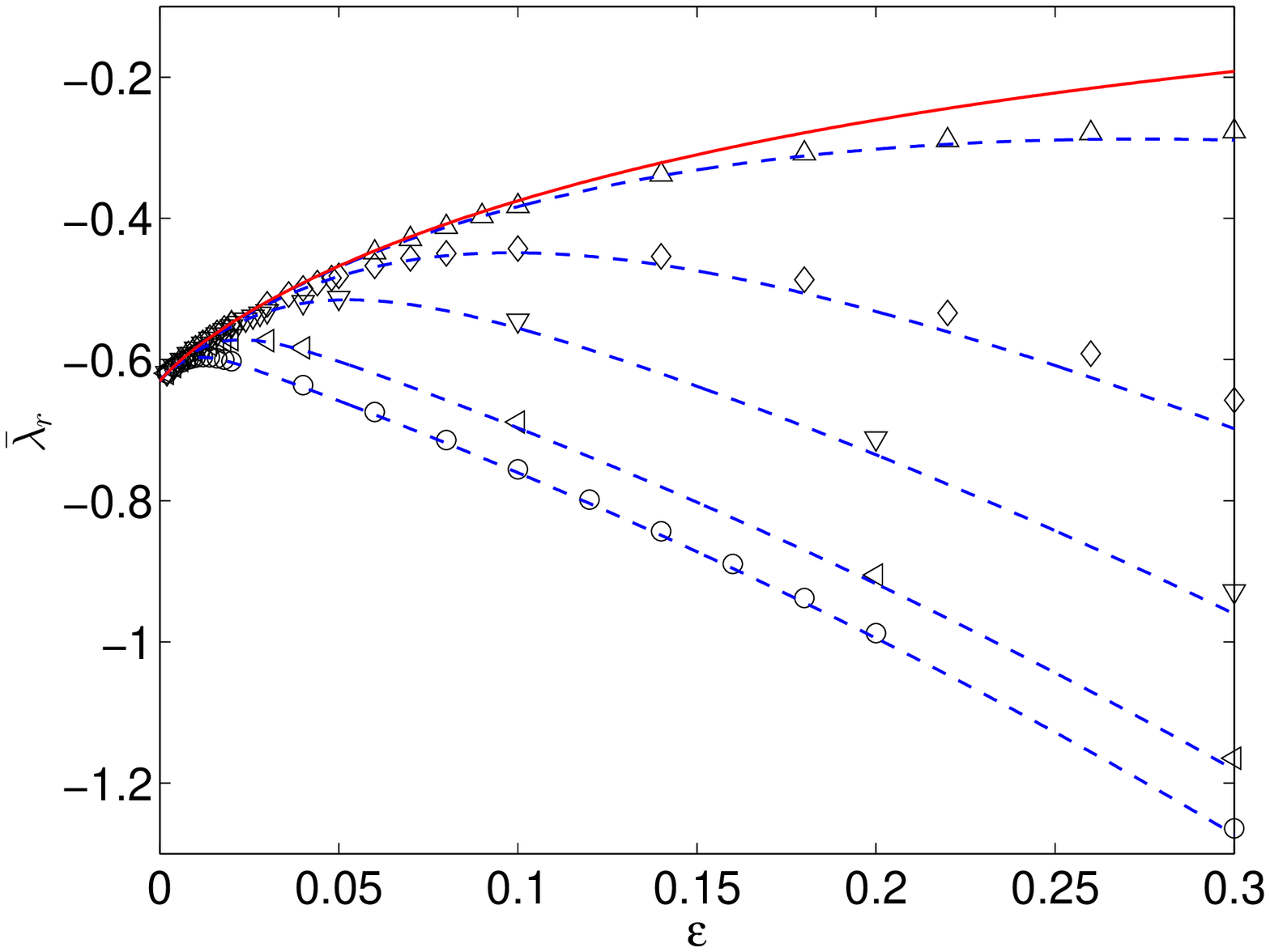}\\
\includegraphics[width=8.6cm]{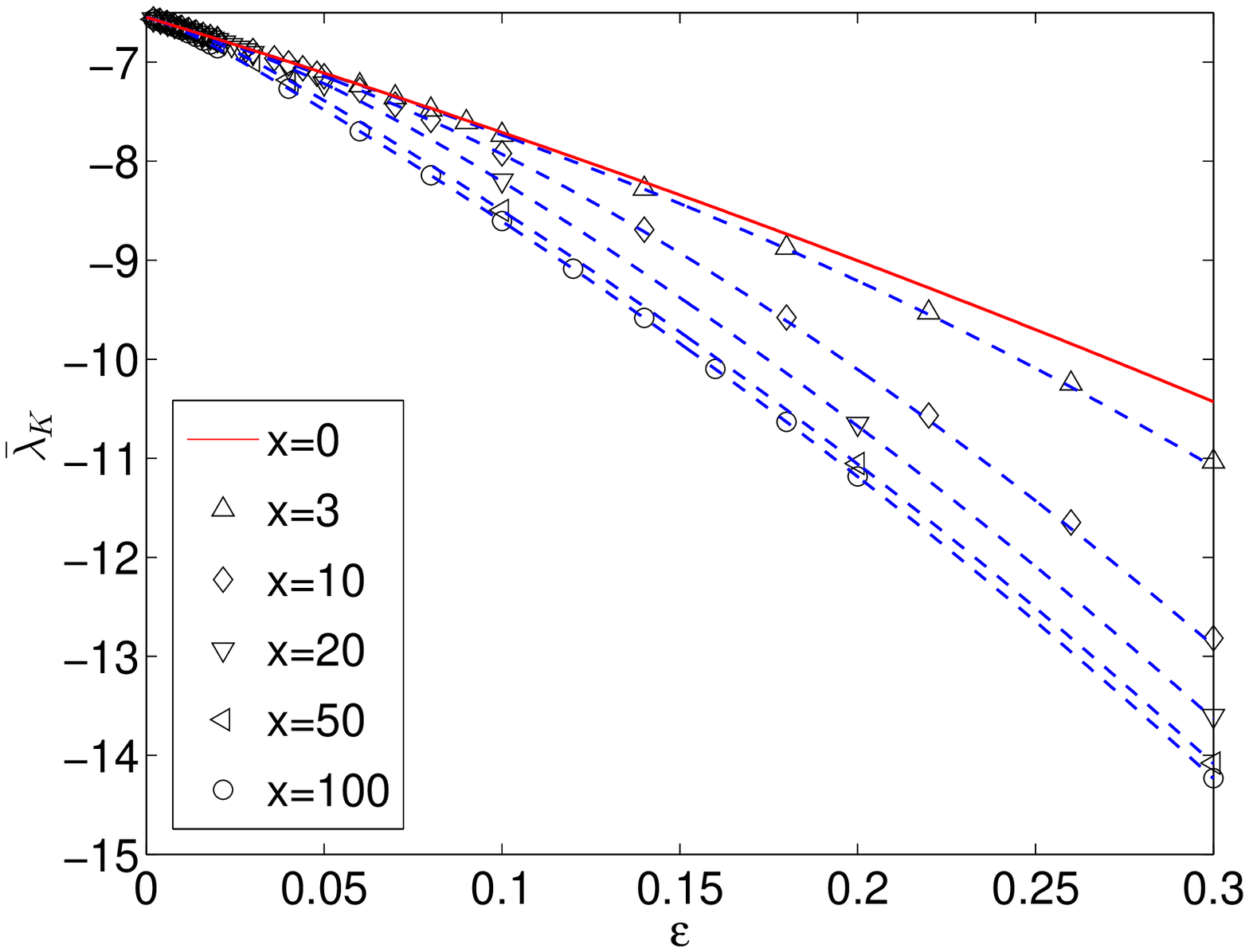} \includegraphics[width=8.6cm]{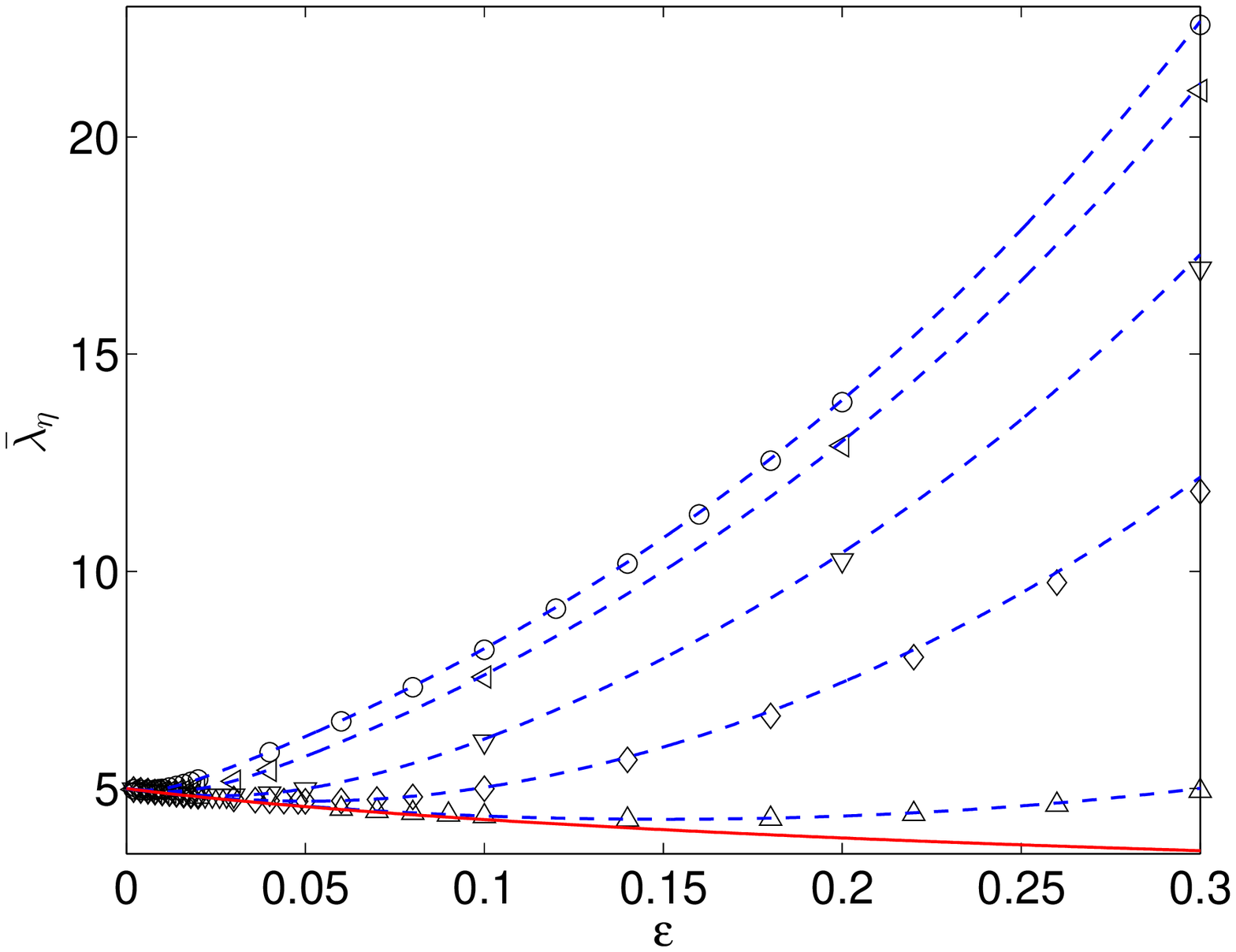}\\
 \caption{The two-particle virial coefficients $\bar{\lambda}(x,\epsilon)$, as functions of the shell thickness $\epsilon$, for suspensions of core-shell particles with $x=0$ (solid line), and $x=3$, $10$, $20$, $50$, $100$ (symbols), in comparison to the effective annulus model, $\bar{\lambda}^A_{\text{eff}}
$ (dashed lines).
Top: the translational and rotational self-diffusion coefficients, $\bar{\lambda}_t$ and $\bar{\lambda}_r$. Bottom: the sedimentation and high-frequency viscosity coefficients, $\bar{\lambda}_K$ and $\bar{\lambda}_{\eta}$.  
}\label{fig1}
\end{figure*}

\begin{table}[h]
 \caption{The annulus model: Two-particle contributions to the short-time translational 
and rotational self-diffusion, $\bar{\lambda}_t^A$ and $\bar{\lambda}_r^A$, sedimentation, 
$\bar{\lambda}_K^A$, and high-frequency viscosity, $\bar{\lambda}_{\eta}^A$.\\}\label{tab1}

\begin{tabular}{lcccc}
\hline
$\delta$&$\bar{\lambda}_{t}^A$ & $\bar{\lambda}_K^A$&$\bar{\lambda}_r^A$&$\bar{\lambda}_{\eta}^A$\\
\hline
\hline
0.000 &  -1.8315 &  -6.5464 &  -0.6305 &   5.0021\\
0.020 &  -1.7821 &  -6.7672 &  -0.5483 &   4.8123\\
0.040 & -1.7397 & -6.9943 & -0.4916 &4.6554\\
0.060 & -1.7015 & -7.2271 & -0.4460 &4.5187\\
0.080 & -1.6664 & -7.4654 & -0.4079 &4.3972 \\
0.100 & -1.6337 & -7.7090 & -0.3752 &4.2877\\
0.120 & -1.6030 & -7.9578 & -0.3466 &4.1881\\
0.140 & -1.5740 & -8.2118 & -0.3214 &4.0970\\
0.160 & -1.5465 & -8.4710 & -0.2990 &4.0132\\
0.180 & -1.5203 & -8.7353 & -0.2789 &3.9357\\
0.200 & -1.4952 & -9.0048 & -0.2608 &3.8639\\
0.220 & -1.4712 & -9.2793 & -0.2444 &3.7970\\
0.240 & -1.4481 & -9.5589 & -0.2294 &3.7347\\
0.260 & -1.4259 & -9.8436 & -0.2158 &3.6765\\
0.280 & -1.4045 &-10.1334 & -0.2033 &3.6220\\
0.300 & -1.3838 &-10.4282 & -0.1919 &3.5708\\
0.320 & -1.3638 &-10.7281 & -0.1813 &3.5228\\
0.340 & -1.3445 &-11.0331 & -0.1716 &3.4777\\
0.360 & -1.3257 &-11.3430 & -0.1626 &3.4352\\
0.380 & -1.3075 &-11.6580 & -0.1542 &3.3952\\
0.400 & -1.2898 &-11.9780 & -0.1465 &3.3574\\
0.420 & -1.2727 &-12.3031 & -0.1392 &3.3217\\
0.440 & -1.2560 &-12.6332 & -0.1325 &3.2880\\
0.460 & -1.2397 &-12.9683 & -0.1262 &3.2560\\
0.480 & -1.2239 &-13.3084 & -0.1203 &3.2258\\
0.500 & -1.2085 &-13.6535 & -0.1148 &3.1972\\
\hline
\end{tabular}
\end{table}

The effective annulus intrinsic viscosity is the same as for a solid sphere,
\bee
\bar{[\eta]}^A(\delta) &=& \frac{5}{2}.
\eee

As discussed in Sec.~\ref{main} and illustrated in Fig.~\ref{fi2}, the effective annulus approximation corresponds to $c=a_{\text{eff}}$, with the effective radius given by Eqs.~\eqref{thesame}-\eqref{r3}. 

To compare the core-shell virial coefficients $\bar{\lambda}(x,\epsilon)$ with the corresponding values for the effective annulus model, we first evaluate the effective radius $a_{\text{eff}}$ and the corresponding effective shell thickness $\epsilon_{\text{eff}}$ from Eqs.~\eqref{r1}-\eqref{r3} and \eqref{epse}, then we evaluate from Table~\ref{tab1} 
the two-particle virial coefficients $\bar{\lambda}^A(\epsilon_{\text{eff}})$ for the annulus model, 
and finally perform the appropriate rescaling of the volume fraction $\phi_{a_{\text{eff}}}=4\pi n a_{\text{eff}}^3/3$  to $\phi_a=4\pi n a^3/3$ (which is used in defining $\bar{\lambda}(x,\epsilon)$ for the core-shell model),
\bee
\bar{\lambda}^A_{\text{eff}}
&\equiv& \bar{\lambda}^A(\epsilon_{\text{eff}})\left(\frac{a_{\text{eff}}}{a}\right)^{\alpha},
\eee
with $\alpha=3$ for $\lambda_{t}$, $\lambda_{K}$, $\lambda_{r}$  and $\alpha=6$ for $\lambda_{\eta}$.
The effective annulus intrinsic viscosity, 
\bee
\bar{[\eta]}^A_{\text{eff}}
 &=& \frac{5}{2} \left(\frac{a_{\text{eff}}}{a}\right)^{3}.
\eee

\subsection{The core-shell particles with thin shells}\label{results}

\begin{figure*}[ht]
\includegraphics[width=7.3cm]{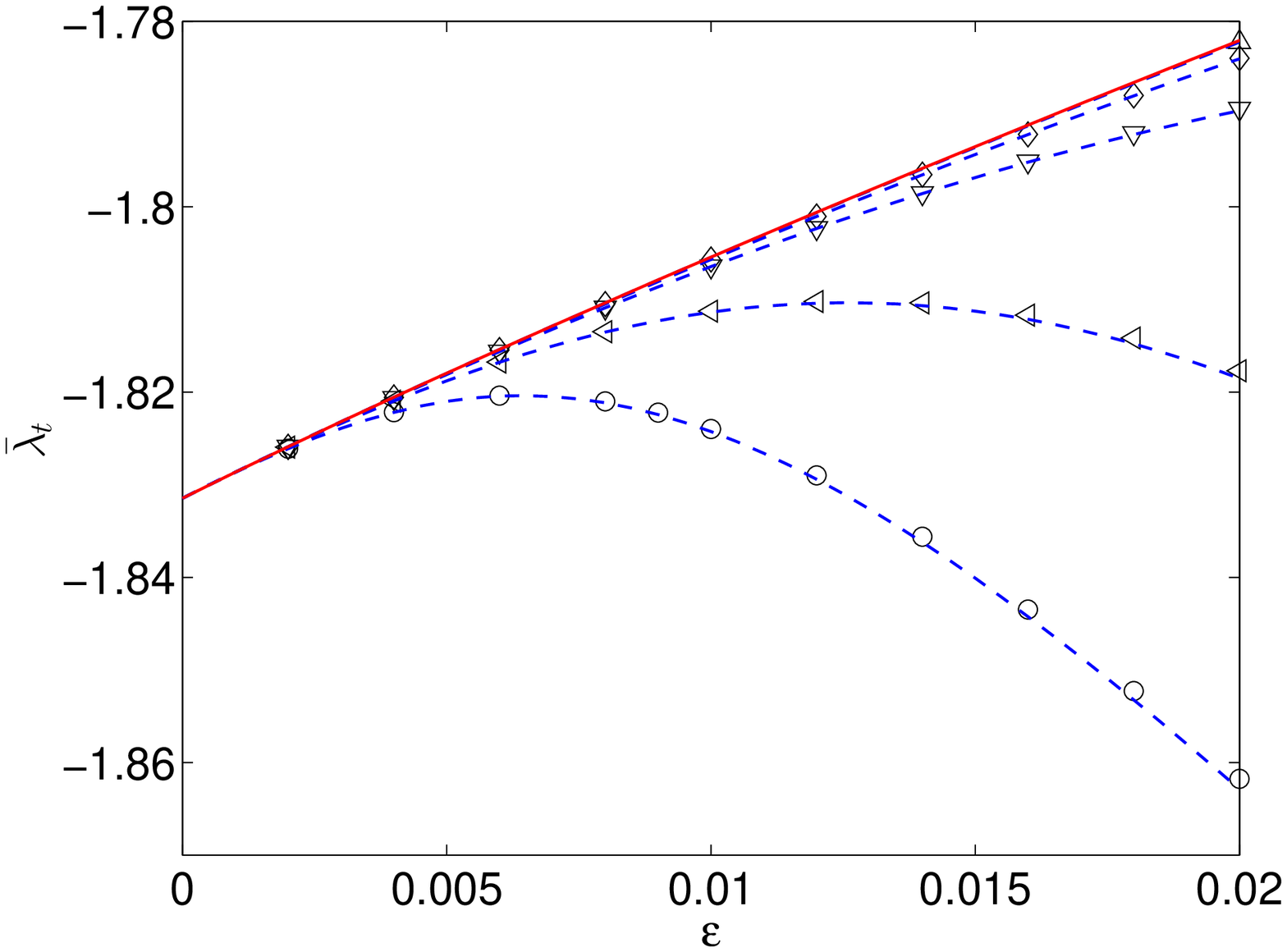} \includegraphics[width=7.3cm]{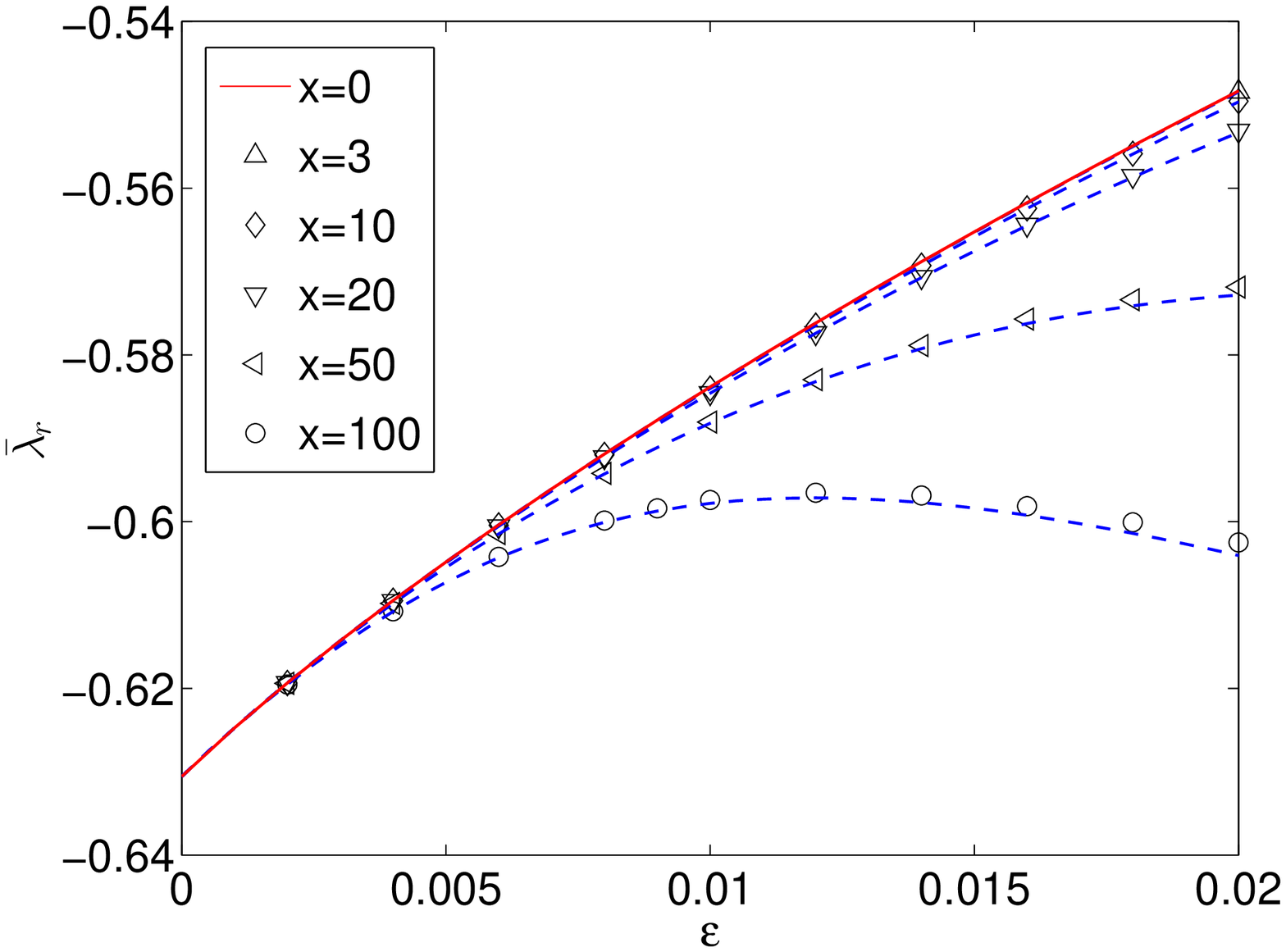}\\
\includegraphics[width=7.3cm]{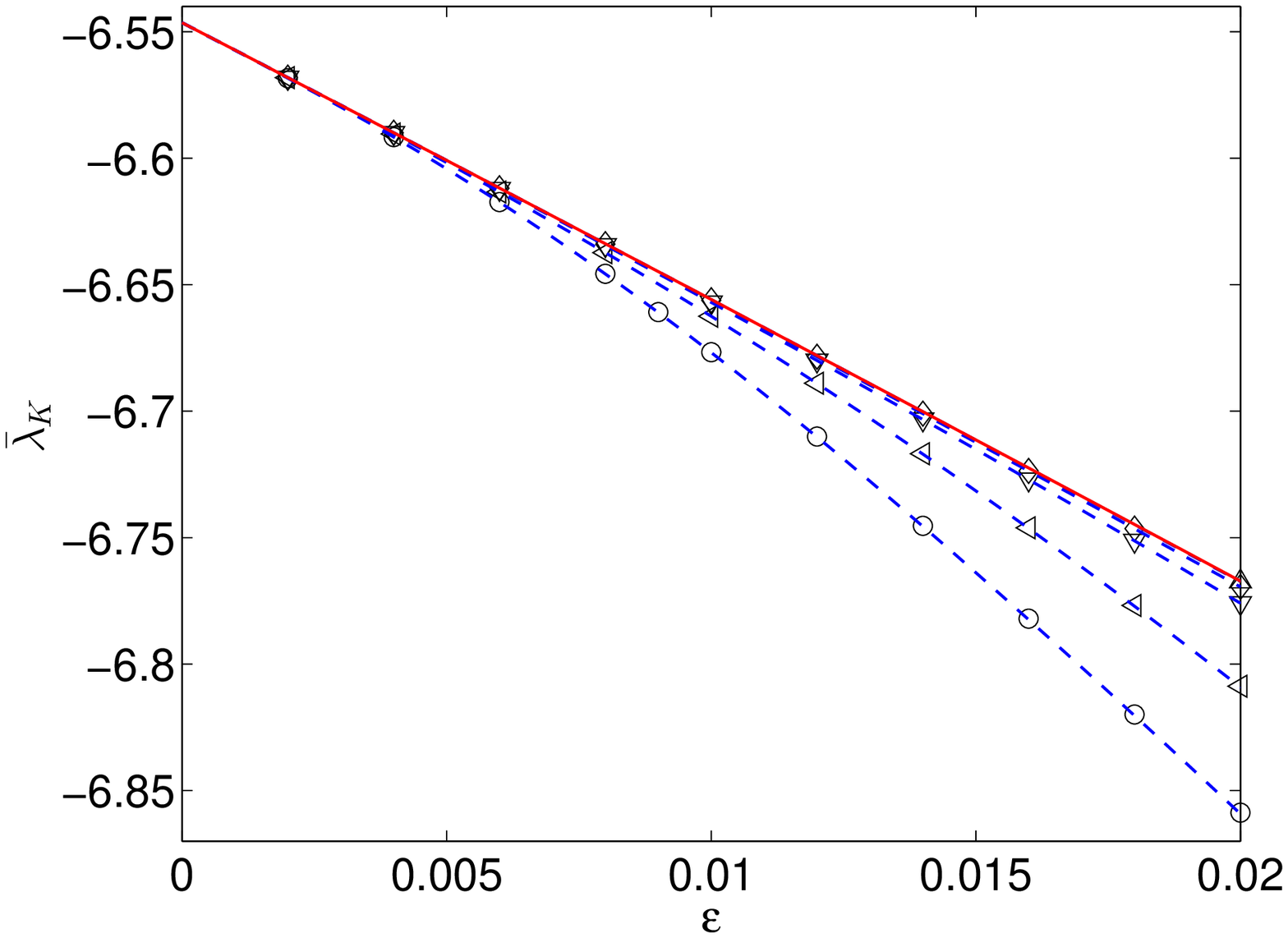} \includegraphics[width=7.3cm]{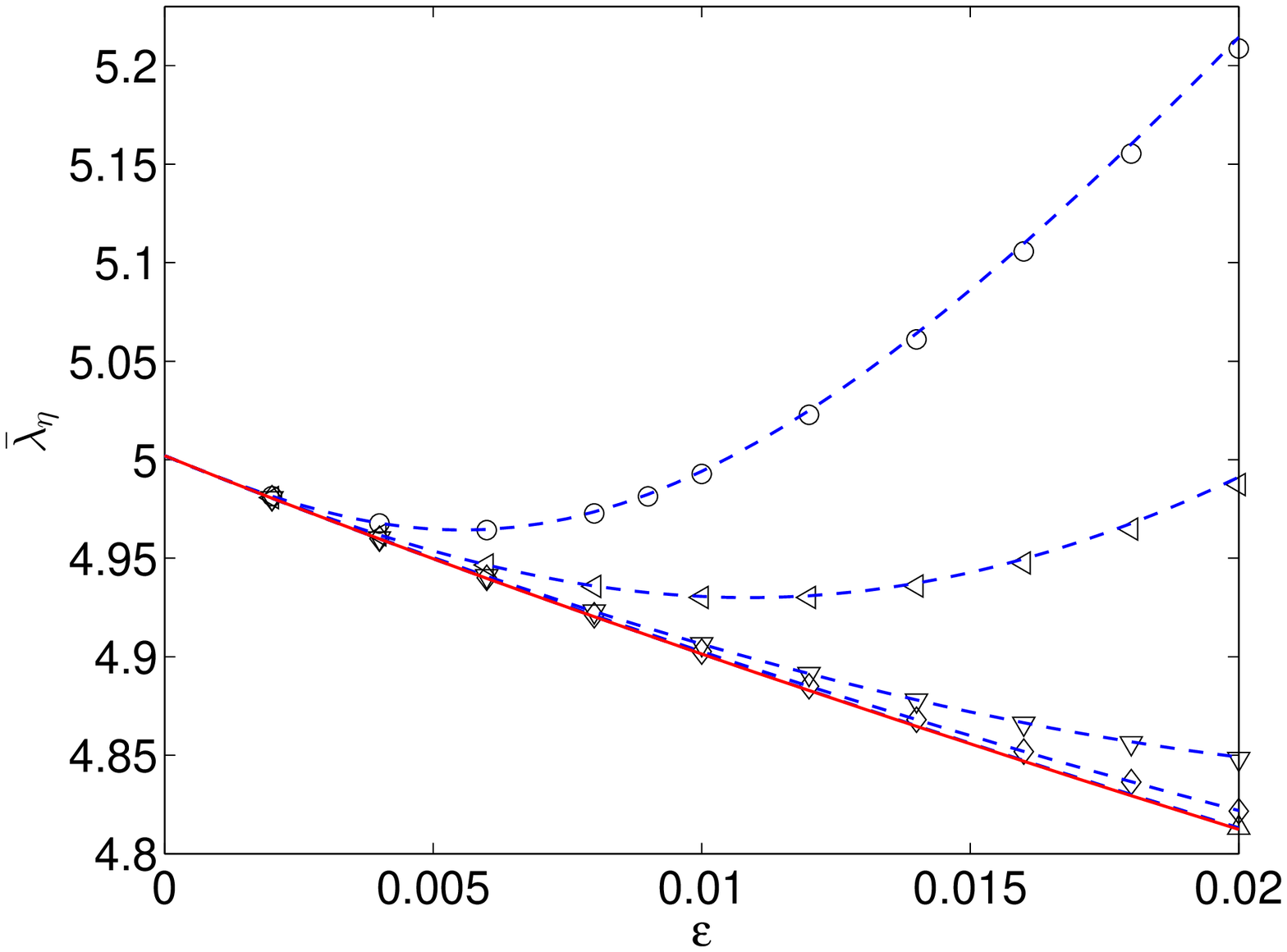}
 \caption{Unlargement of Fig.~\ref{fig1} for very thin shells.
}\label{fi}
\end{figure*}

Based on the previous sections, we now conclude that for core-shell particles with thin shells, 
the virial coefficients $\bar{[\eta]}$ and $\bar{\lambda}$ are well-approximated by the corresponding values following from the effective annulus model,
\bee  
\bar{[\eta]} \!\approx \!\bar{[\eta]}^A_{\text{eff}} \hspace{0.25cm}\mbox{and} \hspace{0.25cm}\bar{\lambda} \!\approx \!\bar{\lambda}^A_{\text{eff}} \hspace{0.25cm}\mbox{for } \epsilon \! \rightarrow \! 0,
 \label{s}
\eee
with the neglected terms scaling as $\sim (x^2\epsilon^4)$.
With the use of the expressions from Appendix~\ref{A}, the two-particle virial coefficients $\bar{\lambda}(x,\epsilon)$ have been evaluated. In Figs.~\ref{fig1} and \ref{fi}, 
 the values of $\bar{\lambda}$ for the core-shell particles are  compared 
with their analogs $\bar{\lambda}^A_{\text{eff}}$ for the effective annulus model. 
It is clear that the effective annulus approximation is very accurate, with the range of validity even wider than it could be expected from the thin-shell  expansion \eqref{thin-shell}.

\section{Conclusions}\label{con}
There are two main conlcusions following from this work. The first one is that
for core-shell particles with thin shells, characterized by large permeabilities, the effective annulus model very  accurately approximates the self-diffusion, sedimentation and viscosity coefficients of core-shell suspensions. This general statement has been 
derived from the asymptotic expansion of the scattering coefficients in the thin-shell limit ($a, \kappa=\mbox{const}, 
\epsilon \rightarrow 0$), and explicitly illustrated by evaluation of the two-particle virial contributions to the transport coefficients, $\bar{\lambda}(x,\epsilon)$, which are shown in Fig. \ref{fig1}. Actually, 
the range of validity of the effective annulus approximation, $\bar{\lambda}= \bar{\lambda}^A_{\text{eff}}[1+{\cal O}(\epsilon^4)]$, with the neglected terms $\sim x^2 \epsilon ^{4}$, extends beyond the case of $x^2\epsilon^4 \lesssim 1$.

\begin{figure}[h]
\includegraphics[width=8.6cm]{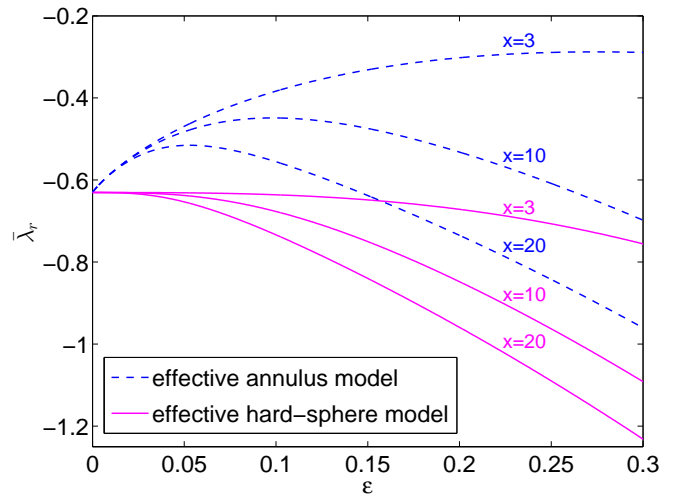} 
 \caption{For very permeable shells, the effective hard-sphere model is not accurate, in contrast to the effective annulus model, compare with the right upper panel of Fig.~\ref{fig1}.  
}\label{fig5}
\end{figure}

The second important outcome is that the effective annulus model for suspensions of core-shell particles with highly permeable shells is much more accurate than just the effective hard-core approximation, in which the particle shells are totally neglected (as completely permeable and overlapping with each other). This effect is explicitly shown 
on the level of two-particle interactions. Large differences between the accurate effective annulus approximation to $\bar{\lambda}_r$, and the corresponding predictions of the imprecise effective hard-sphere model are illustrated in Fig.~\ref{fig5}.

The comparison of the accuracy of both models applied to $\bar{\lambda}(x\!\!=\!\!10,\epsilon\!\!=\!\!0.3)$ is performed in Table \ref{tab2}. For $x\!\lesssim \!10$ and $\epsilon \!\gtrsim \!0.3$, 
the effective hard-core model is not adequate, especially for the rotational self-diffusion. 
\begin{table}[t]
 \caption{Relative accuracy $\Delta \bar{\lambda}$ of  a virial coefficient $\bar{\lambda}$ in the effective annulus and effective hard sphere models, for $\epsilon\!\!=\!\!0.3$ and $x\!\!=\!\!10$.}\label{tab2}

\begin{tabular}{lrrrr}
\hline
&$\Delta \bar{\lambda}_{t}$ & $\Delta \bar{\lambda}_K$&$\Delta \bar{\lambda}_r$&$\Delta \bar{\lambda}_{\eta}$\\
\hline
\hline
effective annulus model&1\% & 0.2\% & 3\% & 2\%\\
effective hard sphere model \;\;&14\% & 13\%\;\;\; & 66\% & 19\%\\
\hline\end{tabular}
\end{table}

The main result of this paper is to derive, in the thin-shell limit ($a, \kappa=\mbox{const}, 
\epsilon \rightarrow 0$), validity of the effective annulus approximation to the core-shell systems, and to demonstrate how to use table \ref{tab1} of the annulus values to determine accurately the core-shell transport coefficients in the semi-dilute regime (and for relatively thin shells), knowing the geometrical and the hydrodynamical radii of a core shell-particle.

\acknowledgments
M.L.E.-J. and E.W. were supported in part by the Polish NCN Grant No. 2011/01/B/ST3/05691.\\

\appendix
\section{Scattering coefficients}\label{ap1}
The single-particle friction operator ${\bf Z}_0$ is expressed in terms of the scattering coefficients by the following relation, 
\begin{widetext}
\begin{eqnarray}
 Z_{0}(ilm\sigma,il'm'\sigma') = \delta_{l l'}\delta_{mm'}\eta(2l\!+\!1) \!
 \left [\!\!
 \begin{array}{ccc}
 \displaystyle \frac{2l(2l\!-\!1)}{l\!+\!1}A_{l0} & 0 & (2l\!-\!1)(2l\!+\!1)A_{l2} \\ \\
 0 &\! l(l\!+\!1)A_{l1} \!& 0 \\ \\
 (2l\!-\!1)(2l\!+\!1)A_{l2} & 0 & \displaystyle \frac{(l\!+\!1)(2l\!+\!1)^{2}(2l\!+\!3)}{2l}B_{l2}
 \end{array}\!\!
 \right ],\!\!\!\!\!\!\!\!\!\!\!\!\!\nonumber \\ \nonumber \\
\end{eqnarray}
\end{widetext}
and 
\bee
\hat{\bf Z}_0 = {\bf Z}_0 - {\bf Z}_0\, \m_0\, {\bf Z}_0,
\label{Zhat}
\eee
with the only non-zero matrix elements of $\m_0$ for $l=1$ and $\sigma=0,1$,
\bee
\mu_{0}(i1m\sigma,i1m'\sigma) &=& [Z_{0}(i1m\sigma,i1m'\sigma)]^{-1}.
\eee
The explicit (lengthy) expressions for the scattering coefficients $A_{l\sigma }(x,a,b)$ for core-shell particles have been derived in Ref.~\cite{Cichocki_Felderhof:09}.
Their asymptotic expansion in the thin-shell limit \eqref{thin-shell} is given in Sec.~\ref{main}. Below, we list the scattering coefficients 
for hard spheres of radius $a$, 
\begin{eqnarray}
&&A_{l0} = \frac{2l+1}{2}a^{2l-1}, \hspace{0.8cm} A_{l1} = a^{2l+1}, \nonumber \\
&&A_{l2} = \frac{2l+3}{2}a^{2l+1}, \hspace{0.8cm} B_{l2} = \frac{2l+1}{2}a^{2l+3}.\hspace{0.9cm}
\end{eqnarray}

\section{Two-particle virial coefficients}\label{A}
In this Appendix, we derive expressions for the two-particle virial coefficients  ${\bar{\lambda}}(x,\epsilon)$ of 
the core-shell suspensions. In the limit of $\kappa$=0 (i.e. $x$=0), the annulus model with the inner radius $c=a$ is recovered.

We introduce dimensionless interparticle distance, $R$=$r/2a$.
The pair distribution function has the form,
\bee
g_0(R) &=& \left\{ \ba{ll} 0& \mbox{ for }\; R\le 1+\epsilon,\\ 1&\mbox{ for }\; R > 1+\epsilon.\ea \right. \label{ggplus}
\eee
which corresponds to the no-overlap condition at the larger shell radius $b=a(1+\epsilon)$. 

The two-particle virial coefficients are evaluated from the following expressions,
\bee
{\bar{\lambda}}_{t}(x,\epsilon)\!\!&=&\!\! 8\int_{1+\epsilon }^{+\infty
}J_{t}(R)R^{2}dR,\label{lat}\\
{\bar{\lambda}}_{K}(x,\epsilon)\!\! &=&\!\!\frac{2}{5a^{3}}A_{12}-\frac{4}{a}%
A_{10}(1+\epsilon )^{2}\nonumber\\
&+&8\int_{1+\epsilon }^{+\infty }J_{K}(R)R^{2}dR, \\
{\bar{\lambda}}_{r}(x,\epsilon)\!\!&=& \!\!8\int_{1+\epsilon }^{+\infty
}J_{r}(R)R^{2}dR,\\
\bar{\lambda}_{\eta}(x,\epsilon)\!\! &=&\!\! \frac{2}{5}\bar{[\eta]}^2 + 24\bar{[\eta]}\int_{1+\epsilon}^{\infty} J_{\eta}(R) R^2 dR, \hspace{0.5cm}
\label{kH}
\eee
where
\bee
J_{t}(R)\!\!&=&\!\!\frac{1}{\mu _{0}^{t}}\mbox{Tr}\,\bm{\mu }_{11}^{tt(2)}(\mathbf{R}),\\
J_{K}(R)\!\!&=&\!\!\frac{1}{\mu _{0}^{t}}\mbox{Tr}\left[ \bm{\mu }_{11}^{tt(2)}(\mathbf{R}%
)+\bm{\mu }_{12}^{tt(2)}(\mathbf{R})-\mathbf{T}_{0}(\mathbf{R)}\right]\!,\hspace{0.6cm}\\
J_{r}(R)\!\!&=&\!\!\frac{1}{\mu _{0}^{r}}\mbox{Tr}\,\bm{\mu }_{11}^{rr(2)}(\mathbf{R}),\\
J_{\eta}(R)\!\! &=&\!\! \frac{1}{5\mu_0^{d} }\left[ {\mu}^{dd(2)}_{11,\alpha \beta \beta \alpha}({\bf R})
+ {\mu}^{dd(2)}_{12,\alpha \beta \beta \alpha}({\bf R}) \right]\!, \hspace{0.5cm}
\label{J}
\eee
with the single particle dipole mobility $\mu_0^{d}$ given by Eqs.~\eqref{invis} and \eqref{mud1}.
In the above expressions, $\mbox{Tr}$ denotes the trace operation, and 
\begin{equation}
\mathbf{T}_{0}(\mathbf{R)=}\frac{\mathbf{1+}\widehat{\mathbf{R}}\widehat{%
\mathbf{R}}}{8\pi \eta_0 R},
\end{equation}
is the Oseen tensor, with
$\widehat{\mathbf{R}}=\mathbf{R/}R$. All the mobility coefficients, the associated functions $J$ and the scattering coefficients $A_{l\sigma}$ in Eqs.~\eqref{lat}-\eqref{J} refer to the core-shell particles.


\end{document}